\documentclass[preprint,11pt]{aastex}

\shortauthors{Winn et al.~2007}
\shorttitle{Super-Neptune HD~149026b}

\begin{document}

%
\def\ltsima{$\; \buildrel < \over \sim \;$}
\def\lsim{\lower.5ex\hbox{\ltsima}}
\def\gtsima{$\; \buildrel > \over \sim \;$}
\def\gsim{\lower.5ex\hbox{\gtsima}}
\def\lam{\lambda=-1\fdg4 \pm 1\fdg1}
%

\bibliographystyle{apj}

\title{
Five New Transits of the Super-Neptune HD~149026b
}

\author{
Joshua N.\ Winn\altaffilmark{1},
Gregory W.\ Henry\altaffilmark{2},
Guillermo Torres\altaffilmark{3},
Matthew J.\ Holman\altaffilmark{3}
}

\altaffiltext{1}{Department of Physics, and Kavli Institute for
  Astrophysics and Space Research, Massachusetts Institute of
  Technology, Cambridge, MA 02139, USA}

\altaffiltext{2}{Center of Excellence in Information Systems,
  Tennessee State University, 3500 John A.\ Merritt Blvd., Box 9501,
  Nashville, TN 37209, USA}

\altaffiltext{3}{Harvard-Smithsonian Center for Astrophysics, 60
  Garden Street, Cambridge, MA 02138, USA}

\begin{abstract}

  We present new photometry of HD~149026 spanning five transits of its
  ``super-Neptune'' planet. In combination with previous data, we
  improve upon the determination of the planet-to-star radius ratio:
  $R_p/R_\star = 0.0491^{+0.0018}_{-0.0005}$. We find the planetary
  radius to be $0.71\pm 0.05$~$R_{\rm Jup}$, in accordance with
  previous theoretical models invoking a high metal abundance for the
  planet. The limiting error is the uncertainty in the stellar
  radius. Although we find agreement among four different ways of
  estimating the stellar radius, the uncertainty remains at 7\%. We
  also present a refined transit ephemeris and a constraint on the
  orbital eccentricity and argument of pericenter, $e\cos\omega =
  -0.0014\pm 0.0012$, based on the measured interval between primary
  and secondary transits.

\end{abstract}


\section{Introduction}

Many clues about the processes of planet formation and evolution have
been discovered by studying the ensemble properties of exoplanets,
such as the ``brown dwarf desert'' (Halbwachs et al.~2000, Marcy \&
Butler 2000) and the tendency for metal-rich stars to have more
detectable planets (Santos et al.~2003, Fischer \& Valenti
2005). However, there are also individual exoplanets whose properties
bear directly on theories of planet formation and evolution. One of
the best examples is the transiting planet HD~149026b (Sato et
al.~2005).

Compared to Saturn, HD~149026b has a similar mass but its radius is
15\% smaller, despite the intense irradiation from its parent star
that should {\it enlarge}\, the radius. Sato et al.~(2005) modeled
HD~149026b as a dense heavy-element core surrounded by a fluid
envelope of solar composition. They found a core mass of
70-80~$M_\oplus$, which is 65-75\% of the total mass of the
planet. This is larger than the canonical core mass of
10-20~$M_\oplus$ that is expected from the core-accretion theory of
planet formation (Mizuno 1980, Pollack et al.~1996). The finding of a
highly metal-enriched composition was confirmed in models by Fortney
et al.~(2006), Ikoma et al.~(2006), Broeg \& Wuchterl~(2007), and
Burrows et al.~(2007). The latter authors dubbed HD~149026b a
``super-Neptune'' because the inferred mass fraction of heavy elements
is similar to that of an ice giant rather than a gas giant.

Interestingly, the parent star has a rather high metallicity
([Fe/H]~$=+0.36$; Sato et al.~2005). The observation of a large core
in such a metal-rich system would seem to support the core-accretion
theory as opposed to coreless alternatives such as gravitational
instability (Boss 1997). However, the larger-than-expected core mass
raises some questions. Why did the growing protoplanet not accrete gas
efficiently?  Or if it did, what happened to its envelope of light
elements?  Many scenarios have been proposed: a collision of two
massive protoplanets (Sato et al.~2005, Ikoma et al.~2006), {\it in
  situ}\, formation in a low-pressure nebula (Broeg \& Wuchterl~2007),
a viscous and evaporating gas disk (Ikoma et al.~2006), and a
separation of gas from planetesimals at the magnetospheric ``X point''
(Sato et al.~2005).

More recently, Harrington et al.~(2007) found that the 8~$\mu$m
brightness temperature of HD~149026b exceeds its expected blackbody
temperature, even if the planet is assumed to absorb all of the
incident stellar radiation. In this sense the planet is anomalously
hot. The high temperature may result from novel atmospheric or
structural properties. Most recently, Torres et al.~(2007) announced
the discovery of a transiting planet, HAT-P-3b, whose measured mass
and radius indicate that it too is highly enriched in heavy elements.

In short, HD~149026b seems to be the harbinger of an entirely new kind
of planet that current models of planet formation, evolution, and
structure cannot accommodate without interesting and possibly exotic
modifications. Because of this situation, it is desirable to improve
the reliability and the precision of estimates of the system
parameters, and especially a key parameter that makes this planet
unusual: its small radius.

One can measure the planetary radius by gathering photometry during
transits, modeling the light curve, and supplementing the model with
external information about the stellar radius.  Previously, Sato et
al.~(2005) analyzed 3 light curves, and Charbonneau et al.~(2006)
added 3 light curves.  In this paper we present another 5 light curves
of comparable or higher quality to the previously published data, and
we simultaneously model all of the data to derive the most precise
planetary, stellar, and orbital parameters that are currently
available. We present our observations and data reduction procedure in
\S~2 and the light-curve modeling procedure in \S~3. We provide the
results in \S~4, along with an extended discussion about the limiting
error: the uncertainty in the stellar radius. The final section
summarizes the results and speculates on future prospects for
improvement.

\section{Observations and Data Reduction}

We used three of the 0.8~m automated photometric telescopes (APTs) at
Fairborn Observatory to measure the transits of HD~149026b that
occurred on UT~2006~April 26, 2006~May~20, 2007~May~3, 2007~June~18,
and 2007~June~21.  We observed the first three transits with the T11
APT and observed the last two transits simultaneously with the T8,
T10, and T11 APTs.  All three telescopes are equipped with two
temperature-stabilized EMI 9124QB photomultiplier tubes for measuring
photon count rates simultaneously through Str\"omgren $b$ and $y$
filters.

On a given night, each telescope automatically acquired brightness
measurements of HD~149026 ($V=8.15$, $B-V=0.61$) and the comparison
star HD~149504 ($V=6.59$, $B-V=0.44$), which was previously
demonstrated to be stable in brightness at the 0.002~mag level or
better (Sato et al.~2005). We also measured the dark count rate and
the sky brightness in the vicinity of each star. We used a diaphragm
of diameter $45\arcsec$ for all the integrations. The integration time
was 20 seconds on the comparison star and 30 seconds on the (fainter)
target star. We computed the magnitude difference for each pair of
target-comparison observations. To increase the signal-to-noise ratio
of each measurement, the differential magnitudes from the $b$ and $y$
pass bands were averaged, resulting in a differential magnitude for a
synthetic $(b+y)/2$ pass band. The typical cadence of the differential
magnitude measurements was 1.4~min for the first 3 transits and
1.2~min for the last 2 transits.

For each raw light curve, we fitted a linear function of time to the
out-of-transit data and divided the data by this function. This was
intended to correct for differential airmass effects and other
systematic errors in the photometry, and also to normalize each light
curve to have unit mean flux outside of the transit. (We also tried
fitting a function of airmass rather than time, but this gave slightly
poorer results.) The final light curves are shown in Fig.~1, and the
data are given in Table~1. The standard deviation of the
out-of-transit data is approximately 0.2\% in all cases, which is
typical for APT observations of bright stars. For additional
information on the telescopes, photometers, observing procedures, data
reduction techniques, and typical photometric precision, see
Henry~(1999) or Eaton, Henry, \& Fekel (2003).  The lower right panel
in Fig.~1 is a composite light curve of all 8 transits observed by the
APTs from Sato et al.\ (2005) and this paper.  The composite light
curve was created by subtracting the mid-transit time from each of the
time stamps, and then averaging into 30 second bins. It is shown here
for display purposes only; the fitting procedure described in the next
section was carried out on the unbinned data.

\begin{figure}[p]
\epsscale{1.}
\plotone{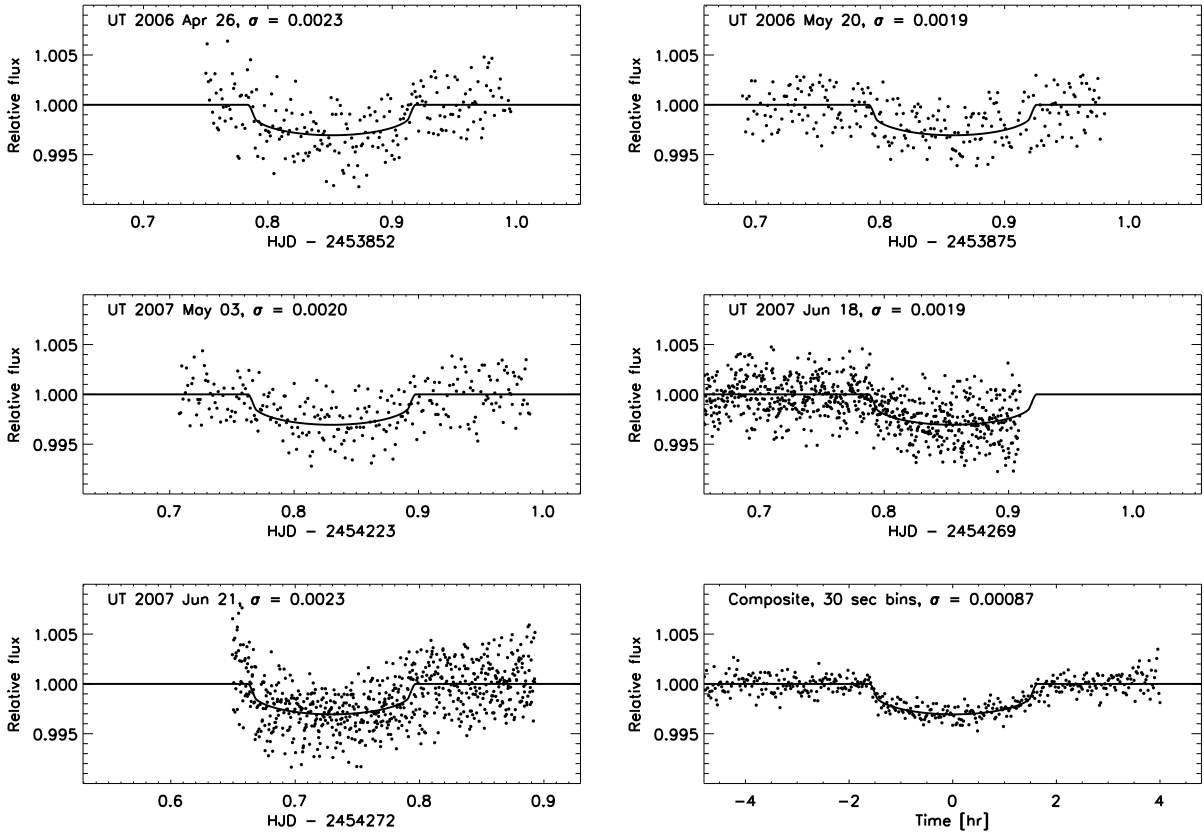}
\caption{
Str\"omgren $(b+y)/2$ photometry of five new transits of HD~149026, 
acquired with three of the 0.8~m APTs at Fairborn Observatory.  The lower 
right panel is a composite light curve created from the five light curves 
reported here as well as the three $(b+y)/2$ light curves previously 
published by Sato et al.~(2005).
\label{fig:1}}
\end{figure}

\section{Determination of System Parameters}

We fitted all of our new photometric data jointly with the three
$(b+y)/2$ light curves presented by Sato et al.~(2005) and the $g$ and
$r$ light curves presented by Charbonneau et al.~(2006).\footnote{We
  did not include the $V$-band light curve of Charbonneau et
  al.~(2006) because of its comparatively large errors and sparse time
  sampling.}  We used a parameterized model based on a two-body
circular orbit. The orbit is specified by the masses of the star and
planet ($M_\star$ and $M_p$), the inclination with respect to the sky
plane ($i$), the orbital period ($P$) and a particular midtransit time
($T_c$). The star and planet are taken to be spheres with radii
$R_\star$ and $R_p$ respectively, and when their sky-projected centers
are within $R_\star+R_p$ of one another we use the Mandel \&
Agol~(2002) formulas to compute the flux decrement due to the partial
blockage of the limb-darkened stellar surface. This is the same code
that has been developed for the Transit Light Curve project (Holman et
al.~2006, Winn et al.~2007). For HD~149026, we assumed the limb
darkening law to be linear, with a coefficient given by Claret~(2000)
for a star of the appropriate temperature and surface gravity. For the
$(b+y)/2$ data we used a coefficient of 0.712, which is the mean of
the tabulated $b$ and $y$ coefficients.

Not all of the parameters listed above can be determined from transit
photometry alone. One set of parameters that can be determined from an
individual light curve is $T_c$, $R_p/R_\star$, $a/R_\star$, and $i$,
where $a$ is the semimajor axis. Our approach was to fix $M_\star$,
$M_p$, and $P$ at previously determined values (thereby fixing $a$
through Kepler's third law), and then fit for $R_p$, $R_\star$, $i$,
and $T_c$. The results for $R_\star$ and $R_p$ are specific to the
choice of $M_\star$, but they scale as $M_\star^{1/3}$ because
$a\propto M_\star^{1/3}$ when the uncertainty in $P$ is negligible, as
it is here. We assumed $M_\star=1.3$~$M_\odot$, following Sato et
al.~(2005), a choice that was subsequently corroborated by our
analysis of the observable stellar properties and the results for
$a/R_\star$ (see \S~4.2).

The fitting statistic was
\begin{equation}
\chi^2 =
\sum_{j=1}^{N_f}
\left[
\frac{f_j({\mathrm{obs}}) - f_j({\mathrm{calc}})}{\sigma_j}
\right]^2
,
\label{eq:chi2}
\end{equation}
where $N_f$ is the number of flux measurements, $f_j$(obs) is the flux
observed at time $j$, $\sigma_j$ controls the weights of the data
points, and $f_j$(calc) is the calculated flux. Experience has shown
that the data weights $\sigma_j$ should account not only for the
single-measurement precision but also the time-correlated (``red'')
noise that afflicts most time-series photometry (see, e.g., Gillon et
al.~2006). The most important timescale in a transit light curve is
the $\sim$10~min duration of the ingress and egress, since the
resolution of ingress and egress is what permits the determination of
$a/R_\star$ and $i$ in addition to $R_p/R_\star$. To assess the noise
on this timescale, we first calculated the standard deviation of the
unbinned out-of-transit data ($\sigma_1$) for each light curve. Then
we averaged the out-of-transit data into 10~min bins consisting of $N$
data points, where $N$ depended on the observing cadence, and
recalculated the standard deviation ($\sigma_N$). In the absence of
red noise, one would expect $\sigma_N = \sigma_1/\sqrt{N}$, but in
practice $\sigma_N$ was larger than $\sigma_1/\sqrt{N}$ by some factor
$\beta$. Therefore, we set the data weights equal to
$\beta~\sigma_1$. The results for $\beta$ ranged from 1.05 to 1.27.

We used a Markov Chain Monte Carlo algorithm to determine the
best-fitting parameter values and confidence intervals. This algorithm
delivers an estimate of the {\it a posteriori} joint probability
distribution for all of the parameters (see Holman et al.~2006 or Winn
et al.~2007 for more details). For each parameter, we took the mode of
the distribution after marginalizing over all other parameters to be
the ``best value.'' We defined the 68\% confidence limits $p_{\rm lo}$
and $p_{\rm hi}$ as the values between which the integrated
probability is 68\%, and for which the two integrals from $p_{\rm
  min}\rightarrow p_{\rm lo}$ and $p_{\rm hi} \rightarrow p_{\rm max}$
were equal.

At first, the preceding computations were performed using free
parameters for both the orbital period and a single midtransit
time. Thus the transits were required to be spaced by integral
multiples of a fixed period. However, we also wanted to measure the
individual transit times, in order to search for variations that might
be indicative of additional bodies in the planetary system (Agol et
al.~2005, Holman \& Murray 2005). To do this, we fixed $R_p$,
$R_\star$, and $i$ at the best values determined in the first step,
and then performed a three-parameter fit of each individual light
curve. The parameters were $T_c$ along with the zero point and slope
of the linear function that was used to correct the out-of-transit
data. (Fixing the values of $R_p$, $R_\star$, and $i$ is justified
because the errors in those parameters are not correlated with the
error in the transit time.) We did this not only for the 5 new light
curves, but also for the 5 previously published light curves, to
provide consistency in the treatment of errors. We then used these
transit-time measurements to refine the estimates of $P$ and $T_c$
(see \S~4.3). For our final results for the photometric parameters
$R_p/R_\star$, $a/R_\star$, and $i$, we reran the MCMC algorithm on
the entire data set, using fixed values of $P$ and $T_c$ from our
refined ephemeris.

\section{Results}

The results of the light-curve analysis are given in Table~2 and
discussed in \S~4.1.  The transit times of the individual light curves
are given in Table~3, and in \S~4.2 we use those times to derive a new
transit ephemeris.  We also use the new ephemeris along with a
previously-measured midpoint of a secondary eclipse to place an upper
bound on one aspect of the orbital eccentricity.  In order to derive
the actual planetary radius (as opposed to the planet-to-star radius
ratio), one must supplement the light-curve analysis with external
information about the stellar radius or mass. In \S~4.3 we investigate
four different methods to estimate the stellar radius, the results of
which are given in Table~4.  The planetary parameters derived from the
light curves and the stellar radius estimates are given in Table~5.

\subsection{Photometric Parameters}

The {\it a posteriori} probability distributions for $R_p/R_\star$,
$a/R_\star$, and $i$ are shown in Fig.~2.  The most well-constrained
of these three basic light curve parameters is the radius ratio,
$R_p/R_\star = 0.0491^{+0.0018}_{-0.0005}$, with a precision of
approximately 2.3\%. The radius ratio is determined largely by the
observed transit depth, which is the smallest among all of the 20
transiting planets known to date.  While there are smaller planets,
such as the Neptune-sized GJ~436b (Gillon et al.~2007), they all orbit
smaller stars, making their radius ratios and transit depths larger
than that of HD~149026.
\begin{figure}[p]
\epsscale{1.0}
\plotone{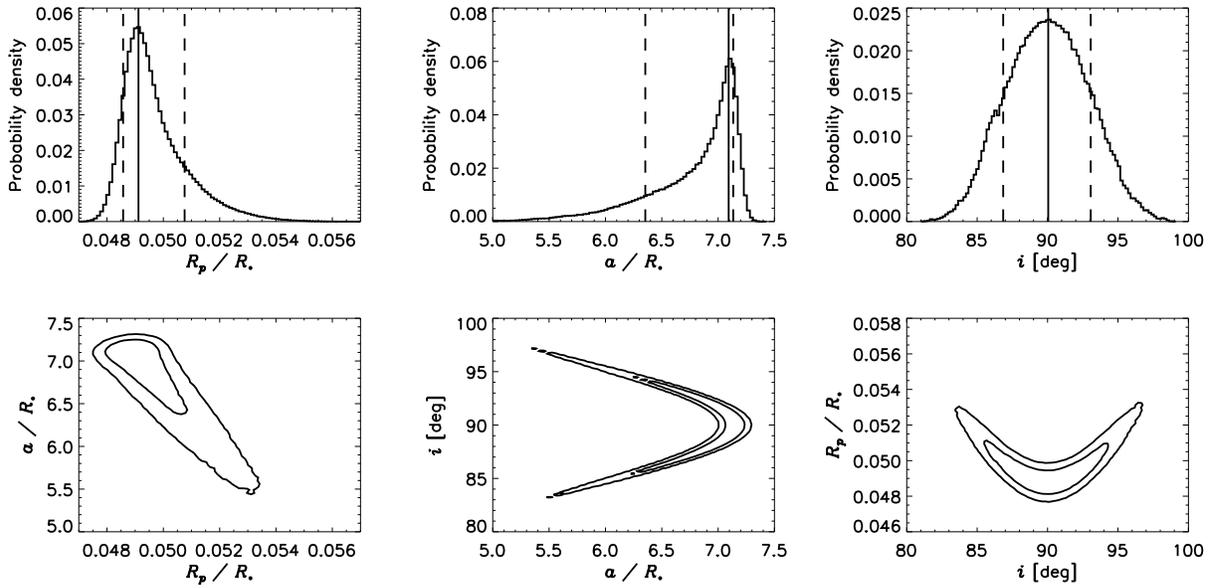}
\caption{ 
Estimated {\it a posteriori} probability distributions from the joint fit to
the transit light curves. The top panels show the
single-variable distributions, in which the mode is marked with a
solid line and the 68\% confidence limits with dashed lines. The
bottom panels show the two-dimensional distributions, in
which the contours mark the 68\% and 95\% confidence limits.
\label{fig:2_01}}
\end{figure}

Less well-constrained are $a/R_\star$ and $i$, the parameters that
depend on the observed durations of the ingress, egress, and total
phases of the transit. Table~2 gives the results for those parameters
as well as the impact parameter and the durations, which can be
derived in terms of $R_p/R_\star$, $a/R_\star$, and $i$.  The data are
consistent with impact parameters ranging from $-0.36$ to $0.36$. As
with any eclipsing binary system, the data cannot distinguish between
positive and negative impact parameters; the probability distributions
are perfectly symmetric about $b=0$ and $i=90\arcdeg$. In this case
the peak probability occurs at the central values $b=0$,
$i=90\arcdeg$. The quantity $a/R_\star$ has a highly asymmetric error
bar. The observed duration of the entire transit event enforces the
upper limit on $a/R_\star$, while the ratio of the ingress (or egress)
duration to the total duration enforces the lower limit on
$a/R_\star$.

Our findings are consistent with the two previous light-curve
analyses, by Sato et al.~(2005) and Charbonneau et al.~(2006),
although our analysis method is different in several ways besides the
use of an expanded dataset. First, we have attempted to account for
time-correlated noise in the photometry, which was neglected in the
previous analyses. Second, unlike the previous authors, we have not
incorporated any {\it a priori}\, constraints on the stellar properties
into our fitting statistic. We made this choice in order to clarify
what information is derived from the light curves themselves; for
example, the previous works did not call attention to the results for
$R_p/R_\star$ even though that parameter is more precisely known than
either $R_p$ or $R_\star$. In addition, our analysis method provides
an estimate of $a/R_\star$ that is independent of any assumptions
about the parent star, except for the very weak dependence on the
chosen limb darkening parameter. This is useful because $a/R_\star$
can be used to determine the stellar mean density (Seager \&
Mallen-Ornelas 2003, Sozzetti et al.~2007, Holman et al.~2007):
\begin{equation}
\rho_\star = \frac{3\pi}{GP^2}\left( \frac{a}{R_\star} \right)^3
 - \rho_p \left( \frac{R_p}{R_\star} \right)^3.
\end{equation}
The last term in this expression may be neglected in this case because
$\rho_p \sim \rho_\star$ and $(R_p/R_\star)^3 \sim 10^{-4}$. Our
independent estimate of $\rho_\star$ is useful in characterizing the
parent star, as described below.

\subsection{The Stellar radius}

To determine the quantity of intrinsic interest, $R_p$, we can
multiply our result for $R_p/R_\star$ by a value of $R_\star$ obtained
by other means. We have investigated four different methods for
determining $R_\star$:

{\it Stefan-Boltzmann Law.}---The bolometric luminosity, effective
temperature, and photospheric radius of HD~149026 are related via
$L_{\rm bol} = 4\pi R_\star^2 \sigma T_{\rm eff}^4$.  We use the {\it
  Hipparcos}\, parallax and apparent magnitude ($\pi = 12.68\pm
0.79$~mas, $V=8.15\pm 0.02$; Perryman et al.~1997) to compute the
absolute $V$ magnitude, apply a bolometric correction of $-0.027\pm
0.014$ (Flower et al.~1996), and use the spectroscopically determined
$T_{\rm eff} = 6147\pm 50~K$~(Sato et al.~2005).  The result is
$R_\star = 1.46\pm 0.10$~$R_\odot$. This is essentially identical to
the value quoted by Sato et al.~(2005) who used the same method.

{\it Spectral Energy Distribution Fit.}---Masana et al.~(2006)
presented an alternative means of estimating the effective temperature
and bolometric correction, using $VJHK$ photometry. They also provided
radius estimates for many nearby stars based on this technique. Using
the {\it Hipparcos}\, parallax and $V$ magnitude along with 2MASS
near-infrared photometry, their result for HD~149026 is $R_\star =
1.515\pm 0.096$~$R_\odot$.

{\it Yonsei-Yale Isochrone Fit.}---Stellar evolutionary models may be
used to estimate the mass, radius, and age of a star with a given
effective temperature, luminosity (or gravity), and metallicity.  We
used the Yonsei-Yale models (Yi et al.~2001, Demarque et al.~2004)
because they are conveniently provided with tools for interpolating
isochrones in both age and metallicity.  For the effective
temperature, we used $T_{\rm eff} = 6160\pm 50$~K, a weighted mean of
the results of Sato et al.~(2005) and Masana et al.~(2006).  We used
the photometric result for $a/R_\star$ as our our proxy for surface
gravity, and we explored the range\footnote{ Sato et al.~(2005)
  reported a metallicity of [Fe/H]~$=+0.36$ with an internal
  uncertainty of $0.05$. To be conservative, we adopted a somewhat
  larger uncertainty of $0.08$, recognizing that different methods for
  determining the metallicity often produce systematic differences of
  this size. } of metallicities [Fe/H]~$=0.36\pm 0.08$. For each
metallicity we considered a range of ages from 0.1 to 14 Gyr, in steps
of 0.1 Gyr. We interpolated the isochrones using a fine mass grid and
compared the points with the measured values of $T_{\rm eff}$ and
$a/R_{\star}$. We computed $\chi^2$ at each point based on the modeled
and observed values of $T_{\rm eff}$, $a/R_\star$, and
metallicity. Then we weighted the points by $\exp(-\chi^2/2)$ and
applied an additional weighting to take into account the density of
stars on each isochrone, assuming a Salpeter initial mass
function. The ``best-fitting'' stellar properties were taken to be the
weighted mean of the properties of all the points. For more details
and other applications of this analysis, see Torres, Winn, \& Holman
(in preparation).  For HD~149026, the results are $M_\star =
1.294_{-0.050}^{+0.060}~M_\odot$, $L_\star =
2.430_{-0.348}^{+0.533}~L_\odot$, and $R_\star =
1.368^{+0.124}_{-0.083}~R_\odot$.  Similar results were obtained when
the spectroscopically determined value of $\log g$ was used instead of
$a/R_\star$.  The theoretical isochrones and the observational
constraints are shown in Fig.~3.
\begin{figure}[p]
\epsscale{0.8}
\plotone{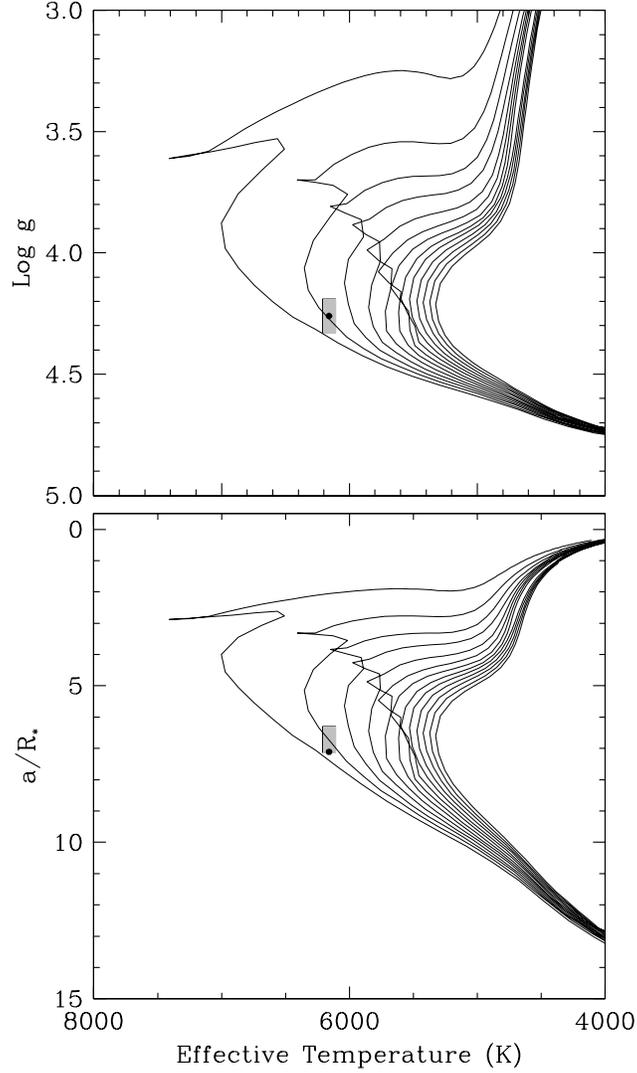}
\caption{ 
Model isochrones from the Yonsei-Yale series by Yi et al.~(2001) and
Demarque et al.~(2004), corresponding to ages of 1--14 Gyr (left to
right), for the measured composition of [Fe/H]~$= +0.36$, along with
the observational constraints. {\bf Top:} The
vertical axis is $\log g$, and the shaded box shows the 1$\sigma$
range based on the spectroscopically determined value of $\log g$.
{\bf Bottom:} The vertical axis is $a/R_\star$, which is proportional to the cube root
of the stellar mean density (see Eq.~2).
The shaded box shows the 1$\sigma$ range based on the photometrically
determined value of $a/R_\star$.
\label{fig:2_02}}
\end{figure}
{\it Kepler's Law with Stellar Mass Prior.}---As mentioned earlier,
the quantity $a/R_\star$ that is determined from the transit
photometry can be used to find $\rho_\star$ (Eq.~2).  With an {\it a
  priori}\, estimate of $M_\star$, one may use $\rho_\star$ to
determine $R_\star$.  Taking $M_\star = 1.30\pm 0.06$~$M_\odot$ based
on the isochrone fit described above, we find $R_\star =
1.35^{+0.17}_{-0.02}$~$R_\odot$.

All of the results for the stellar radius are summarized in Table~4.
They are all consistent with one another at the 1$\sigma$ level, with
a weighted mean of $1.45$~$R_\odot$. However, it must be emphasized
that while the methods are different, they are not wholly
independent. The first two methods both rely on the {\it Hipparcos}\,
parallax, which is the largest source of error in both cases. The
latter two methods both rely on the Yonsei-Yale stellar evolutionary
models. For this reason we cannot say confidently that the uncertainty
in $R_\star$ is any smaller than the uncertainty in each of the
individual measurements, although the mutual agreement is certainly
reassuring. In what follows we adopt the consensus value $R_\star =
1.45\pm 0.10$~$R_\odot$, the same value used in the previous light
curve analyses.

Assuming a Gaussian error distribution for $R_\star$, and the error
distribution for $R_p/R_\star$ obtained from our light curve analysis,
we find the planetary radius to be $R_p = 0.71\pm 0.05$~$R_{\rm Jup}$.
This can be compared to the previously published results of $0.725\pm
0.050$~$R_{\rm Jup}$ (Sato et al.~2005) and $0.726\pm 0.064$~$R_{\rm
  Jup}$ (Charbonneau et al.~2006), keeping in mind our different
method of analysis and treatment of observational errors. The results
are all in agreement. Indeed the differences are smaller than one
would expect from Gaussian statistics, given the quoted error bars,
though we note that 5 of the 10 light curves we fitted were taken from
those previous works. The precision in $R_p$ is not improved because
the limiting error is the uncertainty in $R_\star$, which is
unchanged.

\subsection{Transit Times}

For planning future observations of this system it is important to be
able to predict transit times as precisely as possible. We used all of
the transit times given in Table~3 to calculate a photometric
ephemeris for this system,
\begin{equation}
T_c(E) = T_c(0) + E P,
\label{eq:ephemeris}
\end{equation}
where $T_c$ is the transit midpoint, $E$ is the integral transit
epoch, and $P$ is the orbital period. The linear fit had
$\chi^2/N_{\rm dof} = 0.63$ and $N_{\rm dof} = 9$, suggesting that the
errors quoted in Table~3 have been somewhat overestimated.
The results are:
\begin{eqnarray}
T_c(0) & = & 2454272.7301 \pm 0.0013~{\rm [HJD]} \\
P      & = & 2.8758882 \pm 0.0000061~{\rm days}.
\end{eqnarray}
Our value for the orbital period is in agreement with the previously
published values and is about 25 times more precise. Figure~4 is the
O$-$C (observed minus calculated) diagram for the transit times.

\begin{figure}[p]
\epsscale{1.0}
\plotone{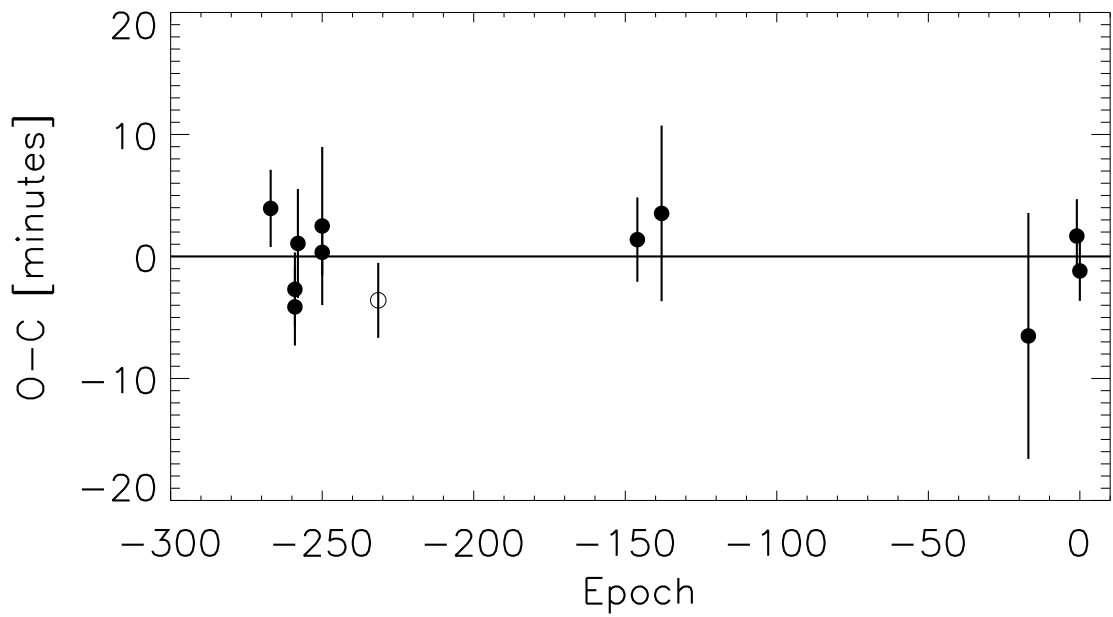}
\caption{ Transit timing residuals for HD~149026b. The calculated times,
  using the ephemeris derived in \S~4.3, have been subtracted from the
  observed times. The filled symbols represent observations of
  transits. The open symbol represents the observation of the secondary
  eclipse by Harrington et al.~(2007).
  The secondary eclipse datum was not used in the fit.
  \label{fig:3}}
\end{figure}

For a circular orbit, successive transits and secondary eclipses
should be spaced by exactly half an orbital period. Recently,
Harrington et al.~(2007) observed a secondary eclipse of HD~149026
with the {\it Spitzer Space Telescope}, allowing the assumption of a
circular orbit to be checked. In the presence of a small but nonzero
orbital eccentricity, the time difference between the midpoint of
secondary eclipse, $T_{\rm sec}$, and the time of transit, $T_{\rm
  tra}$, is
\begin{equation}
T_{\rm sec} - T_{\rm tra} \approx \frac{P}{2}\left(1 + \frac{4}{\pi} e\cos\omega \right),
\end{equation}
where $\omega$ is the argument of pericenter (Kallrath \& Milone 1999,
p.~62). Harrington et al.~(2007) measured the midpoint of a secondary
eclipse to be HJD~$2453606.960 \pm 0.001$, represented by the open
circle in Figure 4. The timing offset of equation (6) is $-3.6\pm
3.1$~minutes, corresponding to $e\cos\omega = -0.0014\pm 0.0012$. The
orbit does indeed appear to be nearly circular, as one would expect
from the dissipative effects of stellar and planetary tidal
interactions.

\section{Discussion and Summary}

We have presented 5 new transit light curves of the exoplanet
HD~149026b and analyzed them along with 5 previously published light
curves. The joint analysis has resulted in much more precise
determinations of the orbital period and transit ephemerides, and also
in a more precise value of the planet-to-star radius ratio. In some
cases, this ratio is of primary interest, such as inferring the
brightness temperature of the planet from the depth of a secondary
eclipse (Harrington et al.~2007), or testing for any wavelength
dependence in the radius ratio as a means of identifying planetary
atmospheric features (see, e.g., Charbonneau et al.~2002).

However, when it comes to understanding the interior structure of the
planet, the quantity of primary interest is $R_p$ itself, and here we
can offer no significant improvement. The limiting error is the 7\%
uncertainty in the stellar radius. This error was not reduced by
acquiring more light curves, although we did find agreement between
the results of 4 different (and intertwined) methods for estimating
the stellar radius using all of the available data. Thus, we leave
unchanged the interpretation of this planet as a being unexpectedly
small for its mass, and likely to be highly enriched in heavy elements
(Sato et al.~2005, Fortney et al.~2006, Ikoma et al.~2006, Burrows et
al.~2006).

Further improvement will depend upon progress in measuring the stellar
radius. Baines et al.~(2007) recently used optical interferometry to
measure the angular diameter of the planet-hosting star HD~189733, and
combined it with the {\it Hipparcos}\, parallax to measure the stellar
radius. For HD~149026, similar observations are not likely to result
in a more precise value of the stellar radius, at least not in the
near future. This is not only because of the 6\% uncertainty in the
parallax, but also because the expected angular diameter is only
$\approx$180~$\mu$as, which is only 7--8 times larger than the
measurement error that was achieved for HD~189733.

Supposing the parallax were known with 10~$\mu$as precision (as one
might hope from a space-based interferometric mission), the error in
the Stefan-Boltzmann method for determining $R_\star$ would be reduced
to 2.7\%. The limiting errors in that case would arise from the
effective temperature and bolometric correction. In the nearer term, a
possible path forward is the continued acquistion of high-quality
transit photometry, in order to improve upon our measurement of
$a/R_\star$ and thereby establish the stellar mean density with
greater precision.  At fixed mean density, $R_\star$ varies as
$M_\star^{1/3}$, and our application of the Yonsei-Yale models to
HD~149026 suggests that the stellar mass has already been pinned down
to within 4.6\%.  If $a/R_\star$ were known exactly, the fractional
error in the stellar radius would be approximately 1.5\% (i.e.,
one-third as large as the fractional error in the stellar mass). In
effect, transit photometry measures $M_\star/R_\star^3$, and the
stellar models generally constrain a different combination of
$M_\star$ and $R_\star$ (see, e.g., Cody \& Sasselov 2002).  We
encourage observers to be persistent in gathering additional seasons
of ground-based photometry and look forward to the results of
space-based photometry for this system.

\acknowledgments We are grateful to the anonymous referee for a
thorough and helpful review of the manuscript. G.W.H.\ acknowledges
support from NSF grant HRD-9706268 and NASA grant NNX06AC14G.  GT
acknowledges partial support for this work from NASA grant NNG04LG89G.

\begin{deluxetable}{lcc}
\tabletypesize{\normalsize}
\tablecaption{Photometry of HD~149026\label{tbl:photometry}}
\tablewidth{0pt}

\tablehead{
\colhead{Heliocentric Julian Date} & 
\colhead{Relative flux} &
\colhead{Uncertainty}
}

\startdata
  2453852.75015 &          1.0031 &          0.0027 \\
  2453852.75115 &          1.0061 &          0.0027 \\
  2453852.75215 &          1.0023 &          0.0027
\enddata 

\tablecomments{The data were obtained with the automatic photometric
  telescopes (APTs) at Fairborn Observatory. Differential magnitudes
  were measured in the Str\"{o}mgren $b$ and $y$ passbands, and the
  $b$ and $y$ results were averaged. The time stamps represent the
  Heliocentric Julian Date at the time of mid-exposure. The
  uncertainties include the ``red noise'' correction described in
  \S~3. We intend for this Table to appear in entirety in the
  electronic version of the journal. An excerpt is shown here to
  illustrate its format.}

\end{deluxetable}

\begin{deluxetable}{llll}
\tabletypesize{\normalsize}
\tablecaption{HD~149026: Transit Light Curve Parameters\label{tbl:params}}
\tablewidth{0pt}

\tablehead{
  \colhead{Parameter} & \colhead{Value} & \colhead{68\% Upper Limit} & \colhead{68\% Lower Limit}
}

\startdata
$R_p / R_\star$               & $ 0.0491 $ & $+0.0018$ & $-0.0005$ \\
$a/R_\star$                   & $ 7.11   $ & $+0.03  $ & $-0.81$   \\
$i$~[deg]                     & $ 90.0   $ & $+3.1   $ & $-3.1$   \\
$b \equiv a\cos i/R_\star$    & $ 0.00   $ & $+0.36  $ & $-0.36$    \\
Total transit duration\tablenotemark{a} & $ 3.254  $ & $+0.057 $ & $-0.028$     \\
Ingress or egress duration\tablenotemark{b} & $ 0.153  $ & $+0.052 $ & $-0.002$
\enddata

\tablenotetext{a}{Defined as the time between first and fourth
  contacts (i.e., between the moments when the projected planetary and
  stellar disks are externally tangent).}

\tablenotetext{b}{Defined as the time between first and second
  contacts (i.e., the duration over which the projected planetary disk
  crosses the stellar limb, from external tangency to internal
  tangency). In our model, the ingress and egress durations are
  equal.}

\tablecomments{Results of fitting ten light curves: three light curves
  [$(b+y)/2$] from Sato et al.~(2005); two light curves ($g$ and $r$)
  from Charbonneau et al.~(2006); and five light curves [$(b+y)/2$]
  from this work. Not all of the parameters are independent. One may
  regard $R_p/R_\star$, $a/R_\star$, and $i$ as the basic parameters
  from which the other results in this table may be derived.}

\end{deluxetable}

\begin{deluxetable}{lccc}
\tabletypesize{\normalsize}
\tablecaption{HD~149026: Midtransit times\label{tbl:times}}
\tablewidth{0pt}

\tablehead{
\colhead{Telescope}   & \colhead{Epoch} & \colhead{Mid-transit time} & \colhead{Uncertainty} \\
\colhead{}            & \colhead{$E$}   & \colhead{[HJD]}            & \colhead{[days]}      
}
\startdata
T11 0.8m APT           & $-267$ & $2453504.8707$ & $ 0.0022 $ \\
T11 0.8m APT           & $-259$ & $2453527.8732$ & $ 0.0021 $ \\
FLWO 1.2m              & $-259$ & $2453527.8722$ & $ 0.0022 $ \\
T8, T10, T11 0.8m APTs & $-258$ & $2453530.7517$ & $ 0.0031 $ \\
FLWO 1.2m              & $-250$ & $2453553.7583$ & $ 0.0013 $ \\
FLWO 1.2m              & $-250$ & $2453553.7598$ & $ 0.0045 $ \\
T11 0.8m APT           & $-146$ & $2453852.8514$ & $ 0.0024 $ \\
T11 0.8m APT           & $-138$ & $2453875.8600$ & $ 0.0050 $ \\
T11 0.8m APT           &  $-17$ & $2454223.8355$ & $ 0.0070 $ \\
T8, T10, T11 0.8m APTs &   $-1$ & $2454269.8554$ & $ 0.0021 $ \\
T8, T10, T11 0.8m APTs &    $0$ & $2454272.7293$ & $ 0.0017 $
\enddata

\tablecomments{Based on these measurements, we derived a transit
  ephemeris $T_c(E) = T_c(0) + EP$ with $T_c(0) =
  2454272.7301(13)$~[HJD] and $P = 2.8758882(61)$~days, where the
  numbers in parentheses indicate the 1$\sigma$ uncertainty in the
  final two digits.}

\end{deluxetable}

\begin{deluxetable}{lll}
\tabletypesize{\normalsize}
\tablecaption{HD~149026: Stellar Radius\label{tbl:star}}
\tablewidth{0pt}

\tablehead{
\colhead{Radius} & \colhead{Method} & \colhead{Reference}
}

\startdata
$1.46 \pm 0.10            $ & Stefan-Boltzmann Law                        &  1,3 \\
$1.515\pm 0.096           $ & Spectral Energy Distribution Fit            &  2   \\
$1.368^{+0.124}_{-0.083}  $ & Yonsei-Yale Isochrone Fit                   &  3   \\
$1.35^{+0.17}_{-0.02}     $ & Kepler's Law with Stellar Mass Prior\tablenotemark{a}  &  3
\enddata

\tablenotetext{a}{Using $M_\star = 1.30\pm 0.06$, based on the
  Yonsei-Yale isochrone fit to $a/R_\star$ and $T_{\rm eff}$.}

\tablecomments{References: (1) Sato et al.~2005; (2) Masana et
  al.~2006; (3) This work.}

\end{deluxetable}

\begin{deluxetable}{lll}
\tabletypesize{\normalsize}
\tablecaption{HD~149026: Planetary Parameters\label{tbl:planet}}
\tablewidth{0pt}

\tablehead{
\colhead{Parameter} & \colhead{Value} & \colhead{Method}
}

\startdata
$M_p$~[$M_{\rm Jup}$]      & $0.36 \pm 0.03           $  &  Spectroscopic orbit\tablenotemark{a}                         \\
$R_p$~[$R_{\rm Jup}$]      & $0.71 \pm 0.05           $  &  $R_p/R_\star$ from light curves and $R_\star = 1.45\pm 0.10$ \\
$\log g_p$~[cgs]           & $3.357^{+0.008}_{-0.130} $  &  Light curve and spectroscopic orbit\tablenotemark{b} \\
$\rho_p$~[g~cm$^{-3}$]     & $1.25\pm 0.28            $  &  $M_p$, $R_p$ given above \\
Semimajor axis, $a$~[AU]   & $0.0432 \pm 0.0006       $  &  Kepler's Law\tablenotemark{c} \\
$e\cos\omega$              & $-0.0014 \pm 0.0012      $  &  Timing of secondary eclipse\tablenotemark{d}
\enddata

\tablenotetext{a}{Using $K=43.3 \pm 1.2$~m~s$^{-1}$, from Sato et
  al.~(2005); $P$ from Table~2; and $M_\star = 1.30\pm 0.06$, based on
  the Yonsei-Yale isochrone fit to $a/R_\star$ and $T_{\rm eff}$.}

\tablenotetext{b}{Using $K=43.3 \pm 1.2$~m~s$^{-1}$, from Sato et
  al.~(2005); $P$ from Table~2; and $i$, $a/R_p$ from the light curve
  analysis. This method is described in detail by Southworth et
  al.~(2007) and Sozzetti et al.~(2007).}

\tablenotetext{c}{Using $P$ from Table~2; and $M_\star = 1.30\pm
  0.06$, based on the Yonsei-Yale isochrone fit to $a/R_\star$ and
  $T_{\rm eff}$.}

\tablenotetext{d}{Using the secondary eclipse time HJD~$= 2453606.960
  \pm 0.001$ from Harrington et al.~(2007) and the ephemeris given in
  Table~2 (after correcting for the 43~s light travel time across the
  orbit).}

\end{deluxetable}

\end{document}